# The $K^+ \to \pi^+ \nu \bar{\nu}$ experiment at CERN

Cristina Biino (for the NA62 collaboration)
*INFN, Sezione di Torino, via P. Giuria 1, 10125 Torino, ITALY*
Monica Pepe (for the NA62 collaboration)
*INFN, Sezione di Perugia, via A. Pascoli, 06123 Perugia, ITALY*

The NA62 experiment, proposed to measure the very rare decay $K^+ \to \pi^+ \nu \bar{\nu}$ at the CERN SPS with 10% accuracy, aims to collect about 80 events in two years of data taking, keeping background contamination lower than 10%. This paper describes the status of the project and the perspectives of the experiment

## 1. INTRODUCTION

The $K^+ \to \pi^+ \nu \bar{\nu}$ decay is a flavor changing neutral current process which proceed through box and purely electroweak penguin diagrams. It is very clean theoretically: short distance dynamics dominates, c-quarks contributions have been evaluated to NNLO order at 5%, and the hadronic matrix elements can be parameterized in terms of the $K^+ \to \pi^0 e^+ \bar{\nu}$ branching ratio that is well known experimentally. For these reasons the $K^+ \to \pi^+ \nu \bar{\nu}$ decay, together with $K_L \to \pi^0 \nu \bar{\nu}$, is extremely sensitive to new physics contributions [1-3]. Moreover, it allows a precise measurement of the CKM parameter $V_{td}$, independent from B oscillation measurements.

The predicted branching ratio is $BR(K^+ \to \pi^+ \nu \bar{\nu})_{theor} = (8.0 \pm 1.1) \cdot 10^{-11}$ [4-7]. The experiment E787/949 at BNL has measured $BR(K^+ \to \pi^+ \nu \bar{\nu}) = (1.73 + 1.15 - 1.05) \cdot 10^{-10}$ [8], compatible with the SM. A 10% accuracy measurement is required to provide a significative test of new physics scenarios.

The NA62 experiment [9], based on the NA48 apparatus, will use the same CERN-SPS beam line which produced the kaon beam for the NA48 experiment and is being designed to reach $10^{-12}$ sensitivity per event, exploiting a decay in flight technique which allows to reach a 10% signal acceptance.

## 2. THE BEAM AND DETECTOR

A $K^+$ beam with a central momentum of 75 GeV/c is produced by a 400 GeV/c intense proton beam ($3.3 \cdot 10^{12}$ protons/pulse) from the SPS impinging on a Be target. The main components are an achromat system made of four dipole magnets and a momentum selecting slit to allow the selection of a narrow momentum ($\Delta P_K / P_K = 1.2\%$), and a second achromat where the beam spectrometer (Gigatracker) is installed. The average rate in the Gigatracker is ~1 GHz, while the one seen by the downstream dectector is ~11 MHz due to kaon decays and accidentals coming from the beam line. The beam is positron free and composed by $K^+$ at about 6.6 %. Assuming 100 days of data taking at 60 % efficiency, about $4.8 \cdot 10^{12}$ $K^+$ decays/year are expected in a 60 m long decay region. The layout of the experiment is shown in Figure 1.

The Gigatracker beam spectrometer measuring kaon momentum and direction, consists of three stations of micro Si pixel detector installed across the second achromat. At least 200 ps time resolution per station is needed to provide a suitable tag of the incident kaon.

The downstream detector contains a magnetic spectrometer made of four low mass straw chambers, each one containing 4 views, placed in vacuum in order to reduce multiple scattering; it will measure precisely the outgoing pion momentum and direction and will act as a veto for other charged particle.

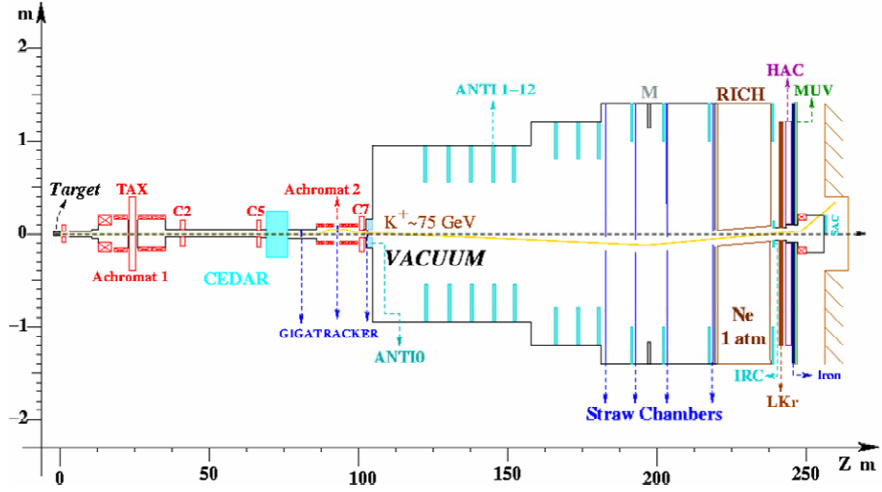

Figure 1: Layout of the NA62 experiment

The particle ID system consists of a CEDAR detector to tag the $K^+$ in the unseparated charged beam and a RICH to identify the decay products. The CEDAR is a differential Cerenkov counter available at CERN beam lines and upgraded to be used in the new beam conditions, namely high intensity. An 18 m long RICH located after the spectrometer and filled with Ne at atmospheric pressure is used to identify pions with momentum greater than 15 GeV/c and will provide also timing information for downstream tracks with a time resolution of 100 ps.

A hermetic combination of calorimeters covering up to 50 mrad and comprising the existing LKr calorimeter, the high performance electromagnetic calorimeter built for NA48 [10], will be used to identify and veto the photons produced in kaon decays. In addition, a muon detector, at the downstream end of the experiment, will be also used as a veto.

The R&D program for the detectors started in 2006 and continues with prototypes integrated in the NA62 set-up and tested during the CERN-NA62 data taking (NA62 is testing Lepton Universality in leptonic decays of charged kaons).

## 3. BACKGROUND REJECTION AND SENSITIVITY

The experimental signature of a $K^+ \rightarrow \pi^+ \nu \bar{\nu}$ decay consists of only one positive track reconstructed in the downstream detectors. The beam and pion spectrometers allow a precise reconstruction of the kinematics, measuring the squared missing mass defined as $m^2_{miss} = (P_K - P_\pi)^2$ under the hypothesis that the final charged particle is a pion. The $m^2_{miss}$ allows a kinematical separation between signal and 92 % of total background, as shown in Figure 2 (left). The backgrounds with the largest branching ratios, $K^+ \rightarrow \pi^+ \pi^0$ and $K^+ \rightarrow \mu^+ \nu_\mu$ decays, force us to define two signal regions where background enters only because of non gaussian tails in the squared missing mass resolution. The remaining 8 % background shown in figure 2 (right) is not kinematically constraint and its rejection must rely on particle identification and the veto system.

Preliminary sensitivity studies on the signal acceptance have been performed using Geant3, Geant4 and Fluka based simulations. A total acceptance of ~14.4 % is obtained, showing that the goal of a 10 % signal acceptance is achievable when taking into account additional losses occurring in real data taking.

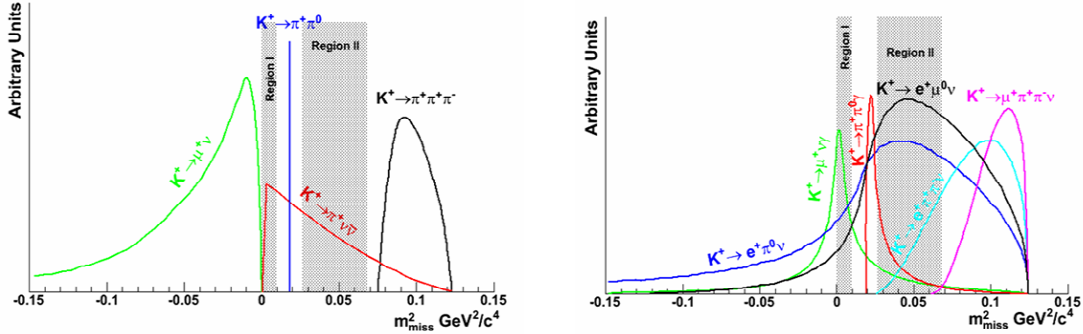

Figure 2: Squared missing mass for 75 GeV/c $K^+$ decays. The signal regions are compared to the kinematically (left) and not kinematically (right) constrained background.

## 4. CONCLUSIONS

Kaon experiments still continue to produce very interesting and exciting results in particle physics. The NA62 Collaboration proposed a very challenging experiment to search for new physics evidences by measuring the BR($K^+ \rightarrow \pi^+ \nu \bar{\nu}$) at the CERN SPS with a 10 % accuracy. A $10^{-12}$ sensitivity per event can be reached using existing infrastructures and part of the NA48 detector. Clearly some new detectors require a sophisticated technology for which an intense R&D program is in progress. NA62 could be ready to start data taking in 2012. The beam will provide a $K^+$ flux ~100 times higher than NA48, giving the opportunity to address many other particle physics phenomena.